# Robust large dimension terahertz cloaking

Dachuan Liang,[1] Jianqiang Gu,[1,*] Jiaguang Han,[1,*] Yuanmu Yang,[1] Shuang Zhang,[2] and Weili Zhang[1,3]

[1]*Center for Terahertz Waves and College of Precision Instrument and Optoelectronics Engineering, Tianjin University, Tianjin 300072, People's Republic of China*

[2]*School of Physics and Astronomy, University of Birmingham, Birmingham, B15 2TT, UK*

[3]*School of Electrical and Computer Engineering, Oklahoma State University, Stillwater, Oklahoma 74078, USA*

Abstract

Invisibility cloaking not only catches the human imagination, but also promises fascinating applications in optics and photonics. By manipulating electromagnetic waves with metamaterials, researchers have been able to realize electromagnetic cloaking in the microwave, terahertz and optical regimes. Nevertheless, the complex design and fabrication process, narrow bandwidth, and high intrinsic losses in the metamaterial-based cloaks have imposed intractable limitations on their realistic applications. Seeking new approaches to overcome these perceived disadvantages is in progress. Here by using uniform sapphire crystal, we demonstrate the first homogenous invisibility cloak functioning at terahertz frequencies. The terahertz invisibility device features a large concealed volume, low loss, and broad bandwidth. In particular, it is capable of hiding objects with a dimension nearly an order of magnitude larger than that of its lithographic counterpart, but without involving complex and time-consuming cleanroom processing. Such homogenous invisibility cloaks may lead to novel terahertz photonic devices potentially used in medicine, radars, and communications.

* Email: gjq@tju.edu.cn; jiaghan@tju.edu.cn

Current interest in the pursuit of exotic electromagnetic properties has been motivated to a large extent by the recent work on invisibility cloaks. Transformation optics based on the invariance of Maxwell's equations has brought the invisibility cloak into reality.[1-4] The experimentally available cloaking designs have been demonstrated in the microwave, terahertz and optical regimes based on artificial metamaterials.[5-10] However, there are two factors that shackle the performance of such cloaking devices, namely the extreme optical parameters and inhomogeneity. As such, it has been difficult to go beyond the limitations in bandwidth, intrinsic loss, small dimension of concealed objects and other undesired features arising from metamaterials.[3] Recently, researchers proposed a novel approach toward macroscopic cloaking based on natural birefringent crystals.[11,12] This straightforward method greatly reduces the complexity in design and fabrication of the metamaterial based cloaks. More importantly, the initial works in the visible regime confirmed that this is a promising way toward the applied macroscopic cloak devices with frequency and incidence angle robustness.[13,14]

Achieving invisibility cloaking in the terahertz regime has recently garnered a great attention due to unique and promising applications of terahertz technology in various fields, such as material characterizations detection for security, molecular sensing, communications, and imaging.[15-17] Particularly, with recent advances in terahertz communications and radars there has been an increasing demand for cloaking devices functioning at terahertz frequencies.[18-20] Recently, the first terahertz carpet cloak was experimentally demonstrated based on microstereolithography (PμSL) fabrication and terahertz time-domain spectroscopy (THz-TDS) measurements.[10,21] The broadband and flexible characteristics of THz-TDS reveal excellent capability in cloaking characterizations. However, fabrication of such terahertz carpet cloaks requires the complex PμSL process, which primarily restricts the dimension of the samples. Here we report a quasi three-dimensional macroscopic cloak functioning in the terahertz regime. The cloak lens is comprised of two sapphire crystal prisms and the experiments were performed in a fiber coupled, broadband THz-TDS system. The cloaking effects were well verified by comparing the cross section profiles of the transverse-magnetic (TM) and transverse-electric (TE) mode terahertz beams that propagated through the invisibility device. This proposed structure has a linear concealed dimension ten times greater than the wavelength, which is the largest cloak device ever observed in the terahertz regime to the best of our knowledge. In particular, our results reveal that the cloaking operates very well from 0.2~1.0 THz and ±10° from the normal incidence angle.

The cloak design is based on recent theoretical and experimental work, which indicated that the carpet cloak can be achieved by specific placement of homogeneous anisotropic crystals with different orientations.[11-14] To cloak the green region as shown in **Figure 1**a, one needs to transform the original space $(x, y)$ shown as the blue triangle in Figure 1b into a new virtual space $(x', y')$, i.e. the orange region in Figure 1a. Then the green region in Figure 1a is not part of the transformation, or in other words, it is optically cloaked. For this purpose we need to take the virtual space in the first quadrant as:

$$x' = x,\ y' = \kappa y + \tau(a - x),\ z' = z, \tag{1}$$

where $\kappa = (\tan\alpha - \tan\beta)/\tan\alpha$, $\tau = \tan\beta$, $\alpha$ and $\beta$ are the bottom angle and cloak angle of the lens, respectively. This new set of coordinates corresponds to a homogeneous constitutive parameter distribution according to Ref. [13], where the permittivity and permeability tensors are:

$$\bar{\bar{\varepsilon}}' = \bar{\bar{\mu}}' = \begin{pmatrix} 1/\kappa & -\tau/\kappa & 0 \\ -\tau/\kappa & \kappa + \tau^2/\kappa & 0 \\ 0 & 0 & 1/\kappa \end{pmatrix}. \tag{2}$$

To simplify the design, we chose the case with TM-polarized terahertz wave in the $x$–$y$ plane and scaled the permittivity tensor to ensure that the ray trajectories are unchanged for $\mu' = 1$. Thus the parameters are totally nonmagnetic and the dielectric tensor is given as:

$$\bar{\bar{\varepsilon}}' = \begin{pmatrix} 1/\kappa^2 & -\tau/\kappa^2 \\ -\tau/\kappa^2 & 1+\tau^2/\kappa^2 \end{pmatrix}. \tag{3}$$

Since the material parameters in equation (3) are spatially invariant and purely dielectric, they can be realized for extraordinary light in nature birefringent materials with special crystal axis orientations. The angle between the optical axis and the $y$ axis is defined as $\theta$ (**Figure 2**a). For a negative uniaxial crystal with ordinary and extraordinary refractive indices $n_o$ and $n_e$, respectively, the relationship between $\theta$, $n_o$, $n_e$ and $H_1$, $H_2$ can be expressed as:

$$H_1/H_2 = 1 - \frac{n_e/n_o}{\sin^2\theta + (n_e/n_o)^2 \cos^2\theta}, \tag{4}$$

where $H_1$ and $H_2$ are the heights of the cloaking region and the whole cloaking lens, respectively. According to equation (4), it is obvious that a larger $n_e/n_o$ leads to a more compact cloak capability. Among the birefringent materials found in the terahertz regime, sapphire crystal is the most suitable material with an ordinary $n_o$ and extraordinary light $n_e$ of 3.068 and 3.406 at 0.5 THz, respectively, giving the highest $n_e/n_o = 1.11$.[16] In addition, the absorption and dispersion of sapphire are both extremely low, which is essential in achieving broadband terahertz cloaking. Based on equation (4), we designed the geometric parameters of our sapphire invisibility device (Figure 2a). The cloaking region has a maximum height $H_2$ = 1.75 mm with a volume of approximately 5% of the whole sample. To illustrate the function of the design, full-wave simulations based on the finite-element method (COMSOL Multiphysics) were carried out for flat surface reflection and wedge reflection from the cloak device with TM and TE polarizations, respectively (Figure 2b, c and d). In the simulations, a polarized Gaussian beam with a central frequency at 0.5 THz entered the lens at normal incidence, i.e. $\varphi_0 = 29.6°$ with respect to the bottom surface (Figure 2a). As shown in Figure 2b and c, for the TM polarization, the profile of the reflected beam from the cloak remains the same as that from a flat surface. However, for the TE mode, the beam was split into two separated parts that propagate in different directions (Figure 2d). In addition, the refraction at the output interface increased the splitting angle significantly, rendering a pronounced contrast between the cloaking and reference beams.

The cloak device was made from two 20-mm-thick high-purity sapphire prisms based on the parameters shown in Figure 2a. All the surfaces were optically polished and the bottom surface was coated with aluminum for a wedge reflection. These two prisms were glued together with mirror symmetry. It has been shown that there is no reflection at the interface between two anisotropic crystals oriented in a mirror symmetry configuration.[22] This is confirmed by a long time scale terahertz scan showing no reflection transient from the joint facet.

The output profiles of the terahertz beam were obtained using an angular resolved reflection THz-TDS (**Figure 3**a). The photoconductive switch-based THz-TDS system was optically gated by 30 fs, 800 nm optical pulses generated from a self-mode-locked Ti:sapphire laser. The TM polarized terahertz radiation emitted from a GaAs transmitter was spatially gathered by a hyper-hemispherical silicon lens and subsequently collimated into a parallel beam before entering the cloaking sample normal to the input facet.[23] The output terahertz wave was received by a TM mode sensitive detector 2 cm away from the output facet of the sample. For comparison, the TE polarized beam profile was chosen as the reference, which exhibits a significant beam splitting (Figure 2d). Since the cloak is polarization sensitive, three polarizers (Microtech Inc.) were added into the system. The Polarizers P1 and P2 were placed in front of the sample to selectively pass the required polarization and P3 was positioned in front of the detector. In the cloaking configuration, the polarization of all three polarizers was parallel to the terahertz electric field. For the reference configuration, P1 and P2 were 45 and 90 degrees oblique from the incident electric field to enable a TE polarized beam. In addition, P3, set at 45 degrees, was combined with the TM polarized detector to receive the refracted TE amplitude signal with a factor of $\sqrt{2}/2$. A schematic diagram of the terahertz cloaking detection and an optical image of the cloak lens are shown in Figure 3b and c, respectively. The terahertz detector scanned the beam profile in parallel to the output facet of the cloak with a 1-mm scan step. As the detector is an incidence angle sensitive receiver,[21] it was rotated at each scan position to ensure the detection of maximum terahertz transient intensity. Subsequently the frequency-dependent terahertz amplitudes at each spatial location were retrieved by Fourier transform of the measured time-domain signals.

In the measurements, the terahertz wave was normally incident ($\varphi_0 = 29.6°$) onto the cloak device. Thanks to the time-domain mechanism of THz-TDS, a broadband cloaking effect from 0.2 to 1 THz can be characterized in a single time-domain measurement at a specific detector location. The Fourier transformed signals were horizontally pixeled with respect to the scan positions, forming the broadband frequency-dependent beam profile maps (**Figure 4**). In Figure 4a, the reflected TM beam from the cloak shows nearly the same profile as that reflected by a flat mirror (Figure 4b). On the other hand, for the TE beam, the detector received two largely separated beam profiles at the left and right side of the cloaking profile (Figure 4c). The experimental result is in good agreement with the simulation prediction illustrated in Figure 2d, where the protruding bottom surface splits the beam into two branches. For a clear comparison of the output beam profiles among the cloaking (Figure 4a), flat surface reflection (Figure 4b) and reference configuration (Figure 4c), the TM and TE measurement results were combined together in Figure 4d. It further verifies that the cloak design unambiguously transforms the bump into a flat surface, realizing a large cloaking volume ($H_2$ = 1.75 mm) underneath the lens. Furthermore, the cloaking effect maintains uniformity over a frequency span of 0.8 THz.

Although a good cloaking performance was demonstrated in Figure 4, a slight split is observed in the cloaked beam profile (Figure 4a). The split, however, is not intrinsic to the cloaking design, but originated from the chamfer located at the bottom corners of the joint facets, which was intentionally added to test the robustness of the cloak against the fabrication latitude. **Figure 5** shows a simulation result on a cloak sample with the chamfer implemented. A narrow dip (dashed line) is clearly reproduced in the output beam profile, but the whole cloaking effect remains quite consistent with that of the experimental results in terms of beam profile and refractive position.

Due the nature of point to point mapping in transformation optics, a perfect invisibility cloak should work at all incidence angles. To investigate the angle dependent cloaking effect, experiments at two other different incidence angles $\varphi_1 = 19.6°$ and $\varphi_2 = 39.6°$ were carried out for both TE and TM polarizations. The results are shown in **Figure 6**a and b and reveal a robust character when compared with that at $\varphi_0 = 29.6°$ (Figure 6c). It is obvious that at both oblique incidence angles, the TM-polarized beam shows no deviation in the profile (the central pattern), while a significant splitting was observed with the TE beam. The corresponding simulation results at these two incidence angles match the measured results extremely well (**Figure 7**), providing further evidence that the birefringent crystalline cloak is indeed a broad angular cloaking device functioning at terahertz frequencies.

Theoretically, cloak designs based on transformation optics is scale independent, which means that the same design can be realized at any electromagnetic frequency. But in fact, terahertz cloak is the least adequately realized among the cloaking works,[5-10,13,14] due primarily to the complication in sample fabrication.[10] In this work, we solved this problem by utilizing a novel design based on birefringence of natural crystals. We have experimentally and numerically demonstrated the first terahertz cloak based on easily processed homogeneous sapphire prisms, which transform a bump mirror into a flat counterpart. The cloak made by two special cut sapphire prisms can hide a triangle area with a height of 1.75 mm, which is record high for terahertz cloaking devices.

Consistent with a more practical and realistic cloaking device functioning in the terahertz regime, we have experimentally verified robustness to operating frequency and incidence angle. Considering these robust qualities and the straightforward fabrication process, the homogenous sapphire invisibility component may open a promising way in novel terahertz cloaking and metamaterial devices.

*Experimental*

*Fiber Coupled THz detector*: The THz-TDS detector was a photoconductive antenna with a 5 μm dipole gap. A 2-meter-long 800 nm single mode fiber was utilized for delivering the ultrafast pumping light onto the antenna. The positive chirp introduced by the fiber was compensated by a pair of gratings shown as a green box in Figure 4a. A collimating lens (f = 11 mm) and a focusing lens (f = 11 mm) were used to focus the pumping light out of the fiber onto the dipole of the terahertz antenna. The output FC/PC port of the fiber, the collimating lens, the focusing lens and the antenna were integrated forming a movable terahertz detector. At the backside of the antenna, a silicon hemispherical lens was attached to efficiently collect and collimate the incidence terahertz wave. A metallic slit with 1 mm width was placed in front of the silicon hemispherical lens to improve the spatial resolution of the scan measurements.

*Acknowledgements*

This work was supported by the U.S. National Science Foundation and the National Science Foundation of China (Grant Nos. 61138001, 61028011, 61007034, and 60977064).

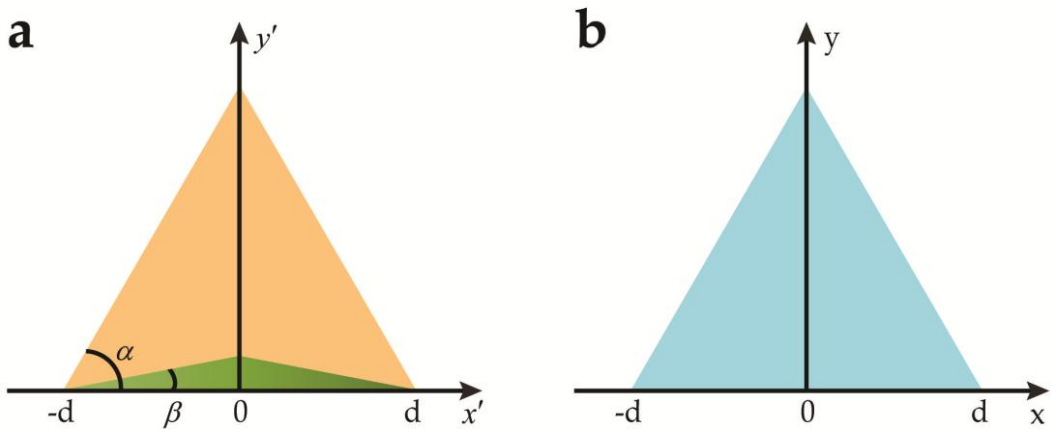

**Figure 1.** Illustrations of the coordinate transformation. A quadrilateral region in (a), which is homogeneous and anisotropic is mapped from a triangular cross-section in the original space (b) made from isotropic material. The cloaked region is illustrated as the green triangle in (a), which conceals objects inside.

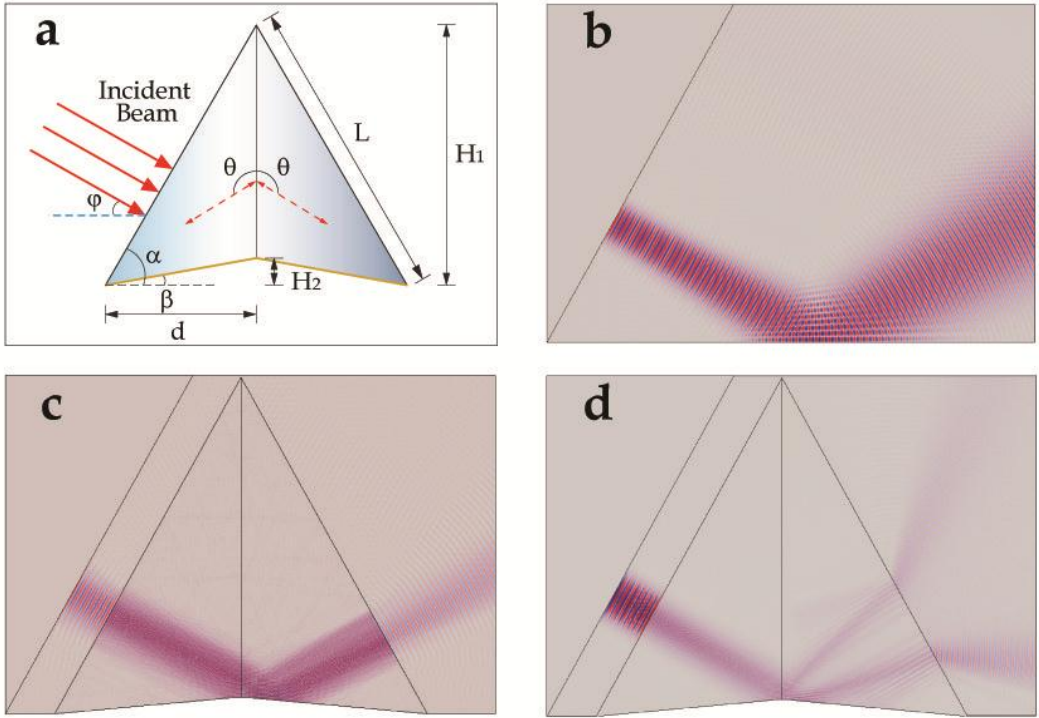

**Figure 2.** (a) Geometric schematics of the cloak lens. The length and height of the lens are $L$ = 40 mm and $H_1$ = 34.78 mm, respectively, forming an bottom angle $\alpha = 60.4°$. The height of the cloaking region is $H_2$ = 1.75 mm, resulting a cloaking angle $\beta = 5°$. The angle $\theta$ can be obtained by diagonalizing the tensor in equation (3): $\theta = 90° - \left|\frac{1}{2}\arctan\frac{2\tau}{\kappa^2 + \tau^2 - 1}\right|$, thus the axis of sapphire was determined as $\theta = 58.5°$. (b)-(d) Simulation results of the terahertz sapphire cloak at normal incidence: (b) flat surface reflection of the TM mode in a homogeneous sapphire, (c) TM mode reflection from the raised bottom of the cloak lens, and (d) TE mode reflection from the raised surface of the cloak lens.

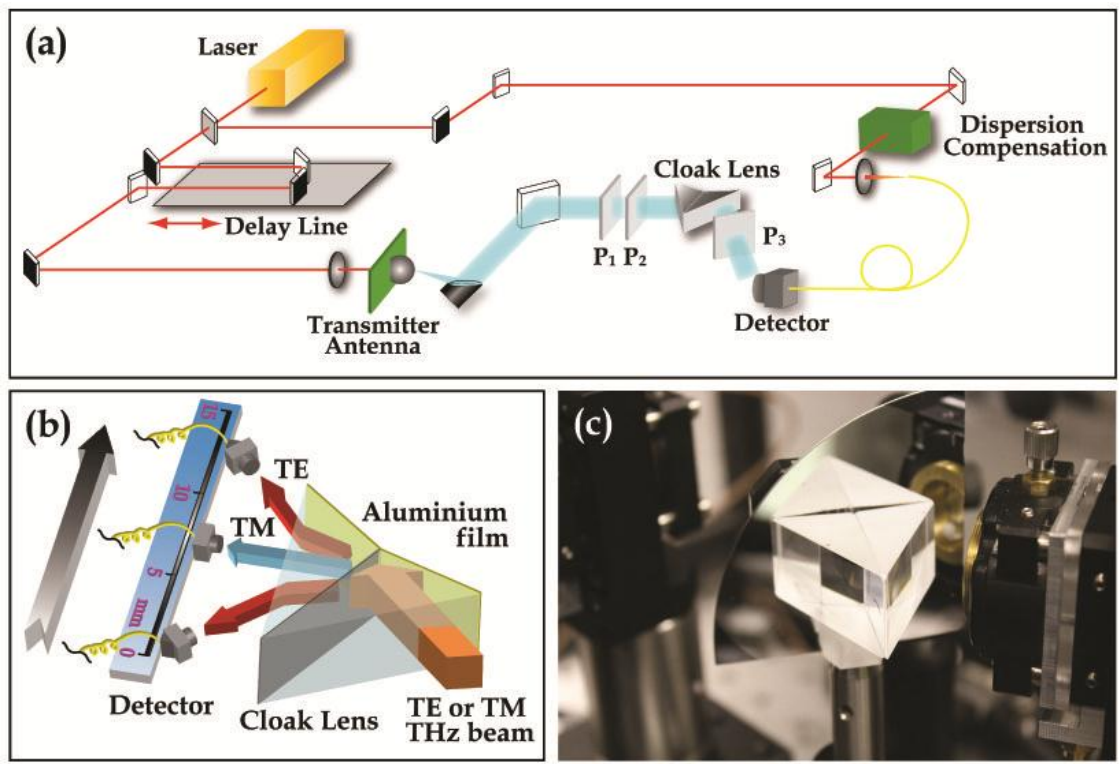

**Figure 3.** Schematic diagram of (a) the experiment setup and (b) the cloaking detection; (c) optical image of the mounted cloak lens.

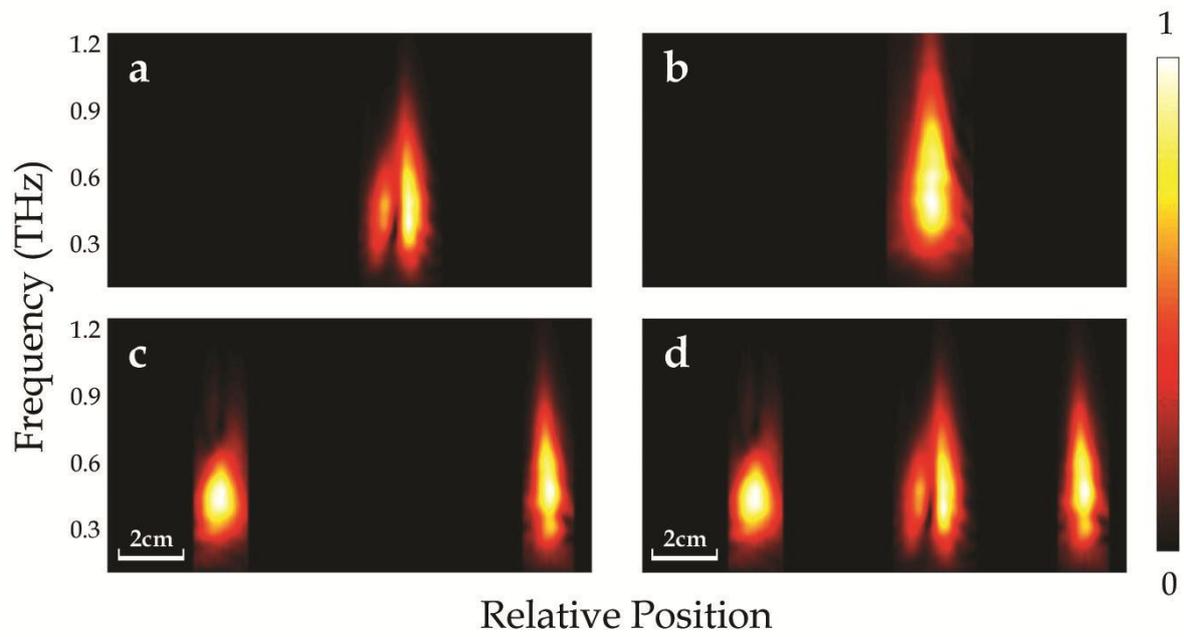

**Figure 4.** Experimental results of the cloaking effect with respect to the relative positions (*x* axis) and the frequency (*y* axis). The color represents the relative spectral amplitude: (a) cloaking, (b) flat surface reflection, (c) reference with the same cloak lens, and (d) combined cloaking and reference profiles.

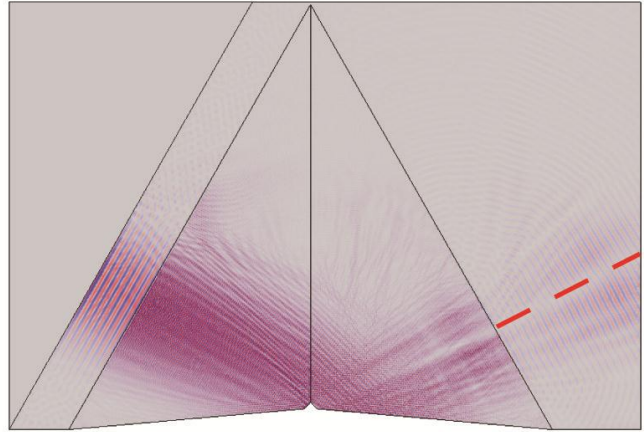

**Figure 5.** Simulation result on the influence of the prism chamfers on the cloaked terahertz beam. A minor split on the output beam was marked by a red dashed line. The chamfers are located right at the center of the cross section of the incident beam, resulting in a split around the center of the output terahertz beam.

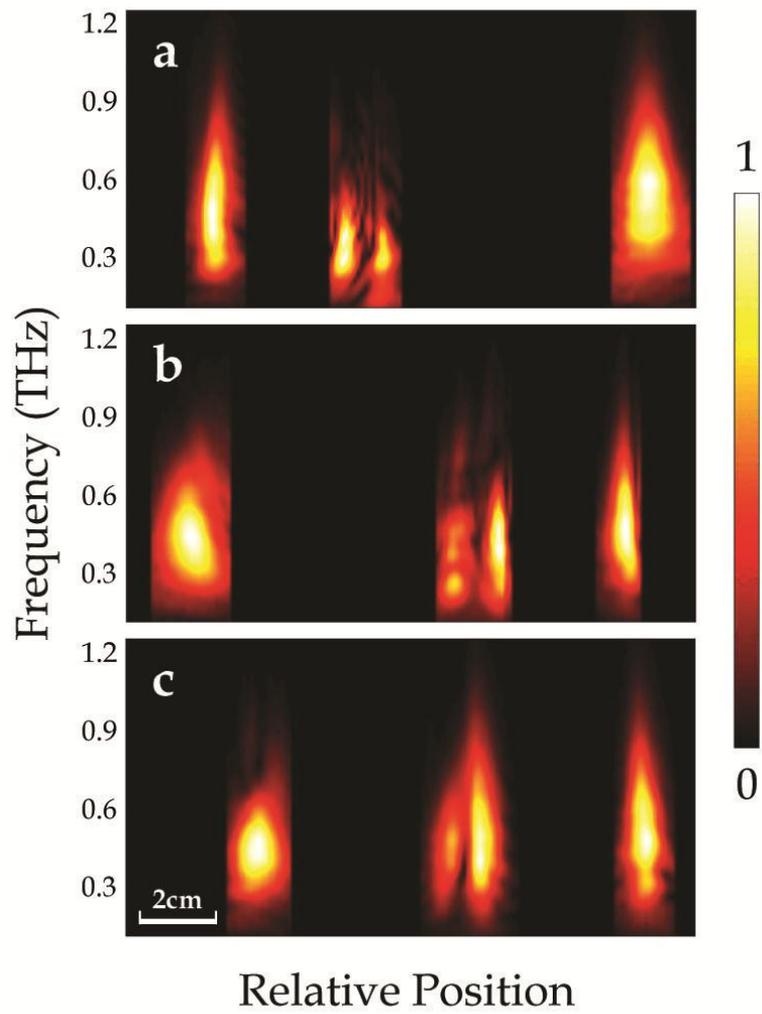

**Figure 6.** TE and TM beam profiles with incidence angles: (a) $\varphi_1 = 19.6°$, (b) $\varphi_2 = 39.6°$, and (c) $\varphi_0 = 29.6°$.

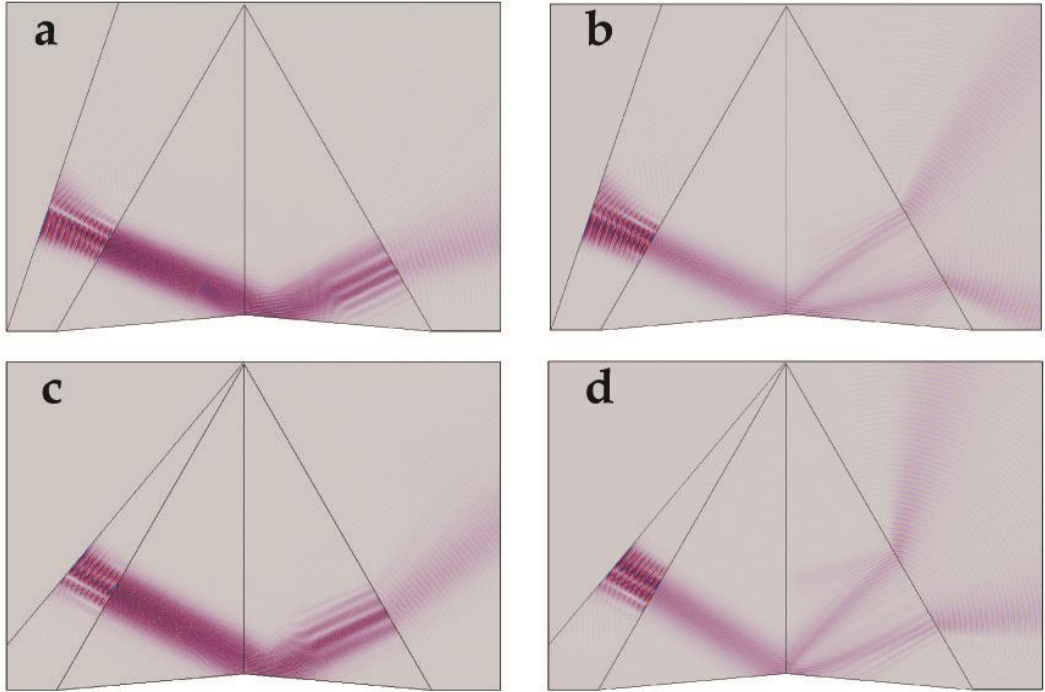

**Figure 7.** Simulated illustrations of the terahertz beam propagating though the cloaking device: (a) TM mode, $\varphi_1 = 19.6°$, (b) TE mode, $\varphi_1 = 19.6°$, (c) TM mode, $\varphi_2 = 39.6°$, and (d) TE mode, $\varphi_2 = 39.6°$.